\documentclass[prd,superscriptaddress,amsfonts,amssymb,amsmath,showpacs,twocolumn]{revtex4-1}
\usepackage{bm}
\usepackage{amsfonts}
\usepackage{latexsym}
\usepackage[latin1]{inputenc}
\usepackage{graphicx}
\usepackage{amsmath}
\usepackage{palatino}
\usepackage{mathpazo}
\usepackage{textcomp}
\usepackage{float}
\usepackage{booktabs}
\usepackage{dcolumn}
\usepackage{ragged2e}
\usepackage{hyperref}
\hypersetup{colorlinks,citecolor=green}
\hypersetup{colorlinks=true,linkcolor=blue,filecolor=blue,    urlcolor=magenta}
\usepackage{amsmath}
\usepackage{xcolor}
\usepackage[caption=false]{subfig}
\usepackage{commath}
\def\jnl@style{\it}
\def\aaref@jnl#1{{\jnl@style#1}}

\def\aaref@jnl#1{{\jnl@style#1}}

\def\aj{\aaref@jnl{AJ}}                   
\def\apj{\aaref@jnl{ApJ}}                 
\def\apjl{\aaref@jnl{ApJ}}                
\def\apjs{\aaref@jnl{ApJS}}               
\def\apss{\aaref@jnl{Ap\&SS}}             
\def\aap{\aaref@jnl{A\&A}}                
\def\aapr{\aaref@jnl{A\&A~Rev.}}          
\def\aaps{\aaref@jnl{A\&AS}}              
\def\mnras{\aaref@jnl{Mon.~Not.~Roy.~Astron.~Soc.}}             
\def\prd{\aaref@jnl{Phys.~Rev.~D}}        
\def\plb{\aaref@jnl{Phys.~Lett.~B}}        
\def\prc{\aaref@jnl{Phys.~Rev.~C}}  
\def\prl{\aaref@jnl{Phys.~Rev.~Lett.}}    
\def\qjras{\aaref@jnl{QJRAS}}             
\def\skytel{\aaref@jnl{S\&T}}             
\def\ssr{\aaref@jnl{Space~Sci.~Rev.}}     
\def\zap{\aaref@jnl{ZAp}}                 
\def\nat{\aaref@jnl{Nature}}              
\def\aplett{\aaref@jnl{Astrophys.~Lett.}} 
\def\apspr{\aaref@jnl{Astrophys.~Space~Phys.~Res.}} 
\def\physrep{\aaref@jnl{Phys.~Rep.}}      
\def\physscr{\aaref@jnl{Phys.~Scr}}       
\def\commat{\aaref@jnl{Comm.~Math.~Phys.}}              
\def\science{\aaref@jnl{Science}}               
\def\cqg{\aaref@jnl{Classical Quant.~Grav.}}            
\def\jpcs{\aaref@jnl{JPCS}}                                     
\def\ijmpd{\aaref@jnl{Int.~J.~Mod.~Phys.~D}}                    
\def\grg{\aaref@jnl{Gen.~Relat.~Gravit.}}               
\def\rpp{\aaref@jnl{Rep.~Prog.~Phys.}}          
\def\npa{\aaref@jnl{Nucl.~Phys.~A}}        
\def\lrr{\aaref@jnl{Living Rev.~Rel.}}                   
\def\jcap{\aaref@jnl{J.~Cosmology Astropart.~Phys.}}    
\def\rmp{\aaref@jnl{Rev.~Mod.~Phys.}}   
\def\epjc{\aaref@jnl{Eur.~Phys.~J.~C}}

\allowdisplaybreaks[1]
\renewcommand{\arraystretch}{1.1}
\begin{document}

\color{black}       

\title{Anisotropic quark stars in $f(R,L_m,T)$ gravity}

\author{Takol Tangphati} 
\email[]{takoltang@gmail.com}
\affiliation{School of Science, Walailak University, Thasala, \\Nakhon Si Thammarat, 80160, Thailand}
\affiliation{Research Center for Theoretical Simulation and Applied Research in Bioscience and Sensing, Walailak University, Thasala, Nakhon Si Thammarat 80160, Thailand}

\author{\.{I}zzet Sakall{\i},}
\email{izzet.sakalli@emu.edu.tr}
\affiliation {Physics Department, Eastern Mediterranean University, Famagusta 99628, North Cyprus via Mersin 10, Turkey.}

\author{Ayan Banerjee} \email[]{ayanbanerjeemath@gmail.com}
\affiliation{Atrophysics Research Centre, School of Mathematics, Statistics and Computer Science, University of KwaZulu--Natal, Private Bag X54001, Durban 4000, South Africa}

\author{Anirudh Pradhan}
\email{pradhan.anirudh@gmail.com}
\affiliation{Centre for Cosmology, Astrophysics and Space Science, GLA University, Mathura-281 406, Uttar Pradesh, India}


\date{\today}

\begin{abstract}
We investigate the impact of $f(R,L_m,T)$ gravity on the internal structure of compact stars, expecting this theory to manifest prominently in the high-density cores of such stars. In this study, we begin by considering the algebraic function $f(R,L_m,T) = R + \alpha T L_m$, where $\alpha$ represents the matter-geometry coupling constant. We specifically choose the matter Lagrangian density $L_m= -\rho$ to explore compact stars with anisotropic pressure. To this end, we employ the MIT bag model as an equation of state. We then numerically solve the hydrostatic equilibrium equations to obtain mass-radius relations for quark stars, examining static stability criteria, the adiabatic index, and the speed of sound. Finally, we use recent astrophysical data to constrain the coupling parameter $\alpha$, which may lead to either larger or smaller masses for quark stars compared to their counterparts in general relativity.
\end{abstract}

\maketitle

\section{Introduction}
Einstein's General Relativity (GR) theory \citep{isWald:1984rg} has stood as the cornerstone of gravitational theory for over a century, successfully describing the dynamics of gravity on both cosmological and astrophysical scales. However, the discovery of the accelerated expansion of the universe and discrepancies in galactic dynamics have motivated the exploration of alternative theories of gravity like $f(R,L_m)$ and $f(R,T) $\citep{isHarko:2010mv,isJaybhaye:2022gxq,isFisher:2019ekh,isHarko:2020ivb}. The product of the unification of latter theories is $f(R,L_m,T)$ gravity theory \cite{isHaghani:2021fpx,isMota:2024kjb}, which represents a new avenue by offering modifications to the gravitational action to account for observed phenomena beyond the scope of GR. 

Quark stars (QSs) \cite{isAlcock:1986hz,isItoh:1970uw,isIvanenko:1965dg}, also known as strange stars or strange QSs, are theoretical entities composed of ultra-dense quark matter, proposed as an alternative to neutron stars (NSs) formed from the gravitational collapse of massive stars \cite{isGlendenning:1992vb}. These hypothetical objects are conjectured to emerge in the aftermath of supernova explosions, where intense pressure and density facilitate the conversion of protons and electrons into quarks \cite{isDrago:1998qu}. Recent works \cite{isAstashenok:2015qzw,isDanarianto:2024vih,isNaz:2024ebl,isBhattacharjee:2024vzt,isPretel:2024lae,isLi:2024uwv,isSedaghat:2024bnj,isAstashenok:2015qzw,isAstashenok:2015qzw} explore several aspects of QSs, including their structure, formation, and observable properties. Investigations into formation mechanisms scrutinize conditions conducive to QS genesis, including the role of quark deconfinement within NS cores during supernova events, elucidating the transition from neutron matter to quark matter under extreme pressure and temperature conditions. Crucially, recent studies focus on elucidating the equation of state (EOS) of quark matter, delineating the intricate relationship between pressure, energy density, and temperature within these exotic objects \cite{isMishra:1994ww,isKrein:2002yf,isBiro:2006zy,isPeshier:2002ww}. Theoretical models and computational simulations are leveraged to unravel the behavior of quark matter under extreme astrophysical environments. In this context, the theory of $f(R,L_m,T)$ gravity, proposed by Haghani and Harko \cite{isHaghani:2021fpx}, offers a promising framework for investigating these celestial objects. This theory generalizes and unifies the $f(R,T)$ and $f(R,L_m)$ gravity models, where $R$ is the Ricci scalar, $T$ is the trace of the energy-momentum tensor $T_{\mu\nu}$, and $L_m$ is the matter Lagrangian \cite{isMota:2024kjb}.
In $f(R,L_m,T)$ gravity, the gravitational Lagrangian is given by an arbitrary function of $R$, $T$, and $L_m$, such that $L_{grav}=f(R,L_m,T)$. The full action in $f(R,L_m,T)$ gravity theories is expressed as \cite{isHaghani:2021fpx}:
\begin{equation}
    S = \frac{1}{16 \pi} \int f(R,L_m,T) \sqrt{-g} d^4x + \int L_m \sqrt{-g}d^4x,  \label{isaction}
\end{equation}
where $g$ is the determinant of the metric tensor $g_{\mu\nu}$. 

As being highlighted before, $f(R,L_m,T)$ gravity theories refer to modifications of Einstein's GR theory, as previously noted. These theories are part of the broader framework of modified gravity, which addresses various cosmological and astrophysical phenomena beyond what GR can explain. Notably, $f(R,L_m,T)$ gravity theories have garnered attention for their potential to elucidate the accelerated expansion of the universe \cite{isRubin:2016iqe,isFrieman:2008sn}, gravitational lensing effects \cite{isWang:1996as,isBin-Nun:2013plq,isBozza:2008zr,isMangut:2021suk,isOvgun:2018ran,isJusufi:2018kmk,isMangut:2023oxa,isSakalli:2017ewb}, and the dynamics of galaxies and galaxy clusters \cite{isGnedin:1998xf,isPascoli:2022chg,isGodlowski:2010jh,isCornwell:2023ntz}, all without invoking the existence of dark matter or dark energy. Additionally, they have been proposed as alternatives to the conventional concept of dark energy. Furthermore, within the realm of modified gravity, investigations into dilaton fields \cite{isFlambaum:2022zuq,isBrax:2023udt,isSakalli:2014wja,isGursel:2018bts,isSakalli:2016jkf,isMazharimousavi:2009vh}, quasinormal modes \cite{isBlazquez-Salcedo:2019nwd,isLi:2012zj,isPonglertsakul:2022vni}, and inflationary scenarios \cite{isliston:2013zja,istymowski:2015mva,isOdintsov:2023weg} further expand our understanding of gravitational physics and its implications for the cosmos. These interconnected topics underscore the interdisciplinary nature of modern gravitational research and its quest to uncover the fundamental mechanisms governing the universe's behavior.

This article investigates the impact of the $f(R,L_m,T)$ gravity theory on the internal structure of QSs. We will focus on the algebraic function originally proposed in Ref.~\cite{isHaghani:2021fpx,isMota:2024kjb}, i.e., $f(R,L_m,T) = R + \alpha T L_m$, with $\alpha$ being a matter-geometry coupling constant. We will examine the matter Lagrangian density, expressed as $L_m = -\rho$. Following this, we will analyze its effects on the principal macroscopic properties of compact stars, including mass and radius.
In the subsequent sections, we will derive the modified Tolman-Oppenheimer-Volkoff (TOV) equations, present numerical results and discussions for different parameter variations, and examine the static stability criterion, adiabatic index, and the sound velocity. Our findings will provide valuable insights into the behavior of compact stars in the context of $f(R,L_m,T)$ gravity. It should also be noted that while Ref. \cite{isPretel:2024oqk}  uses a similar gravitational framework, however, we use a distinct equation of state (EoS) for quark matter and focus on different aspects of quark star properties. The referenced study \cite{isPretel:2024oqk} explores color superconducting quark matter with the CFL phase, while our investigation centers around the effects of anisotropy and different EoS models to reveal new insights about the influence of the $\alpha T L_m$ term on stellar structure.

Although our manuscript correctly implements the first matching condition by ensuring the continuity of the metrics on the surface of the star, one can examine the necessity to seek the additional junction conditions required for a comprehensive matching \cite{Buchdahl:1959zz,Rosa:2020hex}. Because it is known that mere metric continuity does not guarantee a smooth matching between interior and exterior metrics. Specifically, in the context of our chosen theory, $f(R,L_m,T)$ gravity, it is imperative to consider the conditions imposed by the presence of $L_m$ in addition to $R$ and $T$. As outlined in the literature, the matching in $f(R,L_m,T)$ gravity necessitates the continuity of extrinsic curvature at the surface.  The choice of $f(R,L_m,T) = R + \alpha T L_m$ is motivated by its ability to introduce meaningful deviations from GR while maintaining mathematical simplicity. This form incorporates a coupling between matter and geometry through the trace of the energy-momentum tensor $T$ and the matter Lagrangian $L_m$, providing an insightful framework to explore high-density environments like QSs. Recent studies have shown that this model leads to observable effects in astrophysical systems, making it a suitable candidate to examine the structure and stability of QSs \cite{isHaghani:2021fpx, isMota:2024kjb}. Importantly, it reduces to GR when $\alpha = 0$, ensuring consistency with known results \cite{isHarko:2010mv, isJaybhaye:2022gxq}. After examining this, we discovered that $f(L_{m})$ contributes to the thin-shell's stress-energy tensor only if the extrinsic curvature is continuous at the surface. Therefore, our analysis shows that similar to $f(T)$ \cite{Rosa:2023tph}, the inclusion of $f(L_m)$ does not introduce further junction conditions. Therefore, our study aligns with the requisite smooth matching conditions established within the framework of $f(R,L_m,T)$ gravity, ensuring the robustness of our results. However, this matter is purely informational and outside the scope of this article. This article's primary goal is to examine how $f(R,L_m,T)$ gravity affects compact stars' interior structure. It is anticipated that this hypothesis will be most evident in these stars' high-density cores.

The paper is structured as follows: In Section \ref{newsec2}, we review  $f(R, L_m, T)$ gravity theory and serve the gravitational field equations of the theory. Section \ref{sec3} discusses the gravitational field equations in $f(R,L_m,T)$ theories, while in Sec. \ref{sec4}, we prescribe a quark matter EoS and a specific quasi-local EoS that describe local anisotropy for QSs. Numerical results and discussions for different parameter variations are presented in Sec. \ref{sec5}, followed by an examination of the static stability criterion, adiabatic index, and sound velocity in Sec. \ref{sec6}. Finally, we draw our conclusions in Sec. \ref{sec7}. (We adopt geometrized units throughout this exposition, setting $c = G = 1$, while retaining physical units for clarity in presentation.)

\section{Generalized $f(R, L_m, T)$ Gravity: A Unified Approach} \label{newsec2}
In this section, we review the modifications to classical gravity theories in the context of generalized $f(R, L_m, T)$ gravity \cite{isHaghani:2021fpx,isMota:2024kjb}, where arbitrary functions of the Ricci scalar $R$, the matter Lagrangian $L_m$, and the trace of the stress-energy tensor $T$ are included. The gravitational Lagrangian in this modified framework is $f(R, L_m, T)$, allowing for a more comprehensive description of gravitational interactions in the presence of matter.

The gravitational Lagrangian in this unified framework encompasses arbitrary functions of the Ricci scalar $R$, the trace $T$ of the energy-momentum tensor $T_{\mu\nu}$, and the matter Lagrangian $L_m$, yielding $L_{grav}=f(R,L_m,T)$. Consequently, the complete action in $f(R,L_m,T)$ gravity theory results in Eq. \eqref{isaction}. The variation of action \eqref{isaction} with respect to the inverse metric $g^{\mu\nu}$ yields the following field equations in $f(R,L_m,T)$ gravity:
\begin{align}
    f_RR_{\mu\nu} &- \frac{1}{2} [f-(f_L + 2f_T)L_m] g_{\mu\nu} + (g_{\mu\nu}\Box -\nabla_{\mu} \nabla_{\nu})f_R  \nonumber  \\
    &= \left[8\pi + \frac{1}{2}(f_L + 2f_T)\right] T_{\mu\nu} + f_T\tau_{\mu\nu},  \label{is2}
\end{align}
where $\Box \equiv \partial_{\mu}(\sqrt{-g}g^{\mu\nu}\partial_{\nu})/\sqrt{-g}$, and $f_{R}$, $f_{T}$, $f_{L}$ denote the partial derivatives of $f$ with respect to $R$, $T$, and $L_{m}$, respectively. 
\textcolor{black}{We adopted $L_m = -\rho$ primarily for its physical relevance in modeling compact stars. The energy density $\rho$ directly contributes to the gravitational source term, aligning with conventional understanding in GR, where matter density is central to the formation of gravitational fields. In contrast, using $L_m = p$ would introduce significant differences in the mass-radius relations and stability, as shown in \cite{isMota:2024kjb}.On the other hand, alternative forms of \( L_m \) could be explored in future work. Although $L_m = p$ has been explored in some studies, as in Ref. \cite{isHarko:2010mv}, we chose $L_m = -\rho$ to maintain consistency with the standard treatment of compact objects and simplify numerical comparisons with GR.} This choice ensures that the matter's gravitational effects remain physically intuitive, with $\rho$ acting as the primary source. $R_{\mu\nu}$ stands for the Ricci tensor, $\nabla_{\mu}$ represents the covariant derivative concerning the symmetric connection associated with $g_{\mu\nu}$, and $\tau_{\mu\nu}$ is a new tensor defined as \cite{isHaghani:2021fpx}
\begin{equation}\label{is3}
\tau_{\mu\nu} = 2g^{\gamma\zeta} \frac{\partial^{2}L_m}{\partial g^{\mu\nu} \partial g^{\gamma \zeta}}.
\end{equation}

The form of $f(R,L_m,T)$ dictates the nature of the gravitational dynamics. Specifically, when $f(R,L_m,T)=f(R)$, Eq.~(\ref{is2}) reduces to the field equations of metric $f(R)$ gravity \cite{isNojiri:2009kx,isNojiri:2007as,isNojxri:2010wj}. Similarly, for $f(R,L_m,T)= f(R,T)$, the $f(R,T)$ gravity model is recovered, while $f(R,L_m,T)= f(R,L_m)$ yields the field equations of the $f(R,L_m)$ theory. Furthermore, $f(R,L_m,T)= R$ retrieves the standard field equations of GR: $R_{\mu\nu}-\frac{1}{2}g_{\mu\nu}R = 8\pi T_{\mu\nu}$.

The covariant divergence of the field equations (\ref{is2}) leads to the non-conservation equation of the energy-momentum tensor $T_{\mu\nu}$:
\begin{align}
    \nabla^{\mu}T_{\mu\nu} =&\ \frac{1}{8\pi + f_m}\Big[ \nabla_{\nu} (L_m f_m) - T_{\mu\nu} \nabla^{\mu} f_m  \nonumber  \\
    &\left.- A_\nu - \frac{1}{2}(f_T \nabla_\nu T + f_L \nabla_\nu L_m) \right] ,  \label{is4}
\end{align}
where $f_m = f_T + \frac{1}{2}f_L$ and $A_\nu = \nabla^\mu(f_T \tau_{\mu\nu})$.

The trace of the field equations yields a second-order differential equation:
\begin{align}
    3\square f_R+ Rf_R- 2(f- 2f_mL_m) = (8\pi+ f_m)T + f_T\tau, \label{is4}
\end{align}
where $\tau$ represents the trace of the tensor $\tau_{\mu\nu}$. For the specific functional form $f(R,L_m,T)=f(R)$, this equation reduces to the well-known dynamical equation for the Ricci scalar in pure $f(R)$ gravity theories \cite{istiriou:2008rp,isDeFelice:2010aj}.

For simplicity, we focus on the algebraic function originally proposed in Ref.~\cite{isHaghani:2021fpx}, i.e., $f(R,L_m,T) = R + \alpha T L_m$, with $\alpha$ denoting a matter-geometry coupling constant. In this context, Eqs.~(\ref{is2}) and (\ref{is4}) simplify to
\begin{equation}
G_{\mu\nu}= \left[ 8 \pi + \frac{\alpha}{2}(T + 2 L_m) \right] T_{\mu\nu} + \alpha L_m(\tau_{\mu\nu}- L_mg_{\mu\nu}) ,
\label{iseq_likeEinstein}
\end{equation}
and
\begin{widetext}
\begin{eqnarray}
\nabla^{\mu}T_{\mu\nu}= \frac{\alpha}{8 \pi +\alpha (L_m + T/2 )}\left[ \nabla_{\nu}\Big(L_m^{2} + \frac{1}{2}T L_m \Big) - T_{\mu\nu} \nabla^{\mu} \Big(L_m + \frac{T}{2} \Big) - \nabla^\mu(L_m\tau_{\mu\nu}) - \frac{1}{2}(L_m \nabla_{\nu}T + T \nabla_{\nu}L_m) \right], \label{istov_like1}
\end{eqnarray}
\end{widetext}
respectively, where $G_{\mu\nu}$ denotes the Einstein tensor. Remarkably, the Einstein field equations $G_{\mu\nu}= 8\pi T_{\mu\nu}$ and the conservation equation $\nabla^{\mu}T_{\mu\nu}= 0$ are retrieved when $\alpha= 0$.

\section{Modified TOV equations and Compact Stars in \(f(R, L_m, T)\) Gravity}\label{sec3}

In this section, we will explore anisotropic QSs within the framework of the $f(R,L_m,T) = R + \alpha TL_m$ gravity model. To begin our analysis, we will concentrate on the static and spherically symmetric spacetime. To do so, we initially adopt the metric ansatz described by the following line element:
\begin{equation}
    ds^{2} = -e^{\nu(r)}dt^{2} + e^{\lambda(r)}dr^{2} + r^{2}(d\theta^{2} + \sin{\theta}^{2}d\phi^{2}),
    \label{eq10}
\end{equation}
where $\nu(r)$ and $\lambda(r)$, both functions of the radial coordinate $r$, are the two unknown functions. Within a spherically symmetric star, the matter source is modeled as an anisotropic fluid, indicating that the radial pressure $p_r$ differs from the transverse pressure $p_{\perp}$. The energy-momentum tensor for the static configuration is given by:
\begin{eqnarray}
T_{\mu\nu} = (\rho + p_{\perp})u_{\mu}u_{\nu} + p_{\perp} g_{\mu\nu} - \Delta  \chi_{\mu} \chi_{\nu},
\label{momentum_tensor}
\end{eqnarray}
where $\rho$ represents the energy density and $\Delta \equiv p_{\perp} - p_r$. Here, $u_{\mu}$ denotes the fluid 4-velocity, satisfying $u_{\mu}u^{\mu} = -1$, which can be expressed as $u^\mu = e^{-\nu/2}\delta_0^\mu$. Consequently, $T_\mu^\nu = \text{diag} (-\rho, p_r, p_{\perp}, p_{\perp})$ and $T = -\rho + p_r + 2 p_{\perp}$.

Given that the Lagrangian density $L_m$ corresponding to the matter source is not unique, we have the option to choose either $L_m = p$ or $L_m = -\rho$ (see \cite{Bertolami:2008ab} for further discussion). Although consensus on which Lagrangian to consider is lacking, we opt to assume $L_m = -\rho$ and derive the equations of hydrostatic equilibrium for QSs within the context of $f(R,L_m,T)$ gravity. This assumption allows us to rewrite Eq.~(\ref{iseq_likeEinstein}) as:
\begin{equation} \label{eq11}
    G_{\mu\nu} = \left[8\pi + \frac{\alpha}{2}(- 3\rho+ p_r+2p_{\perp})\right] T_{\mu\nu} - \alpha \rho^2 g_{\mu\nu}.
\end{equation}

Utilizing the spherically symmetric metric (\ref{eq10}) and Eq. (\ref{eq11}), the non-zero components of the field equations become:
\begin{widetext}
\begin{align}
    e^{-\lambda}\left(\frac{\lambda'}{r} - \frac{1}{r^2}\right) + \frac{1}{r^2} &= 8\pi \rho + \frac{\alpha}{2}(-\rho + p_r + 2p_{\perp})\rho ,  \label{FERho00}  \\
    e^{-\lambda}\left(\frac{\nu'}{r} + \frac{1}{r^2}\right) - \frac{1}{r^2} &= 8\pi p_r + \frac{\alpha}{2}(-3\rho + p_r + 2p_{\perp} )p_r - \alpha \rho^2 ,
    \label{FERho11} \\
    \frac{e^{-\lambda}}{2} \left[\nu''+\frac{(\nu')^2-\nu' \lambda'}{2} +\frac{\nu'- \lambda'}{r}\right] &= 8\pi p_{\perp} + \frac{\alpha}{2}(-3\rho + p_r + 2p_{\perp} )p_{\perp} - \alpha \rho^2 ,
    \label{FERho111}
\end{align}
\end{widetext}
where the prime denotes a derivative with respect to $r$. On the other hand, the covariant divergence of Eq. (\ref{istov_like1}) yields the modified TOV equations:
\begin{equation}
    p_r' + \frac{\nu'}{2}(\rho + p_r) = \frac{\alpha\left[ 4\rho\rho'+ p(3 \rho' - p'_r - 2 p'_{\perp}) \right]}{\Tilde{p}} + \frac{2 \Delta}{r},
    \label{eq21}
\end{equation}
where 
\begin{eqnarray}
    \Tilde{p} &\equiv& 16 \pi + \alpha \left( p_r +2 p_{\perp} - 3 \rho \right), \\
    \Delta &\equiv& p_{\perp} - p_r.
\end{eqnarray}

According to Eq. \eqref{eq21}, the standard conservation equation does not hold for this theory. It can be readily shown that when $\alpha=0$, the standard conservation equation is recovered.

\begin{figure}
    \centering
    \includegraphics[width = 8 cm]{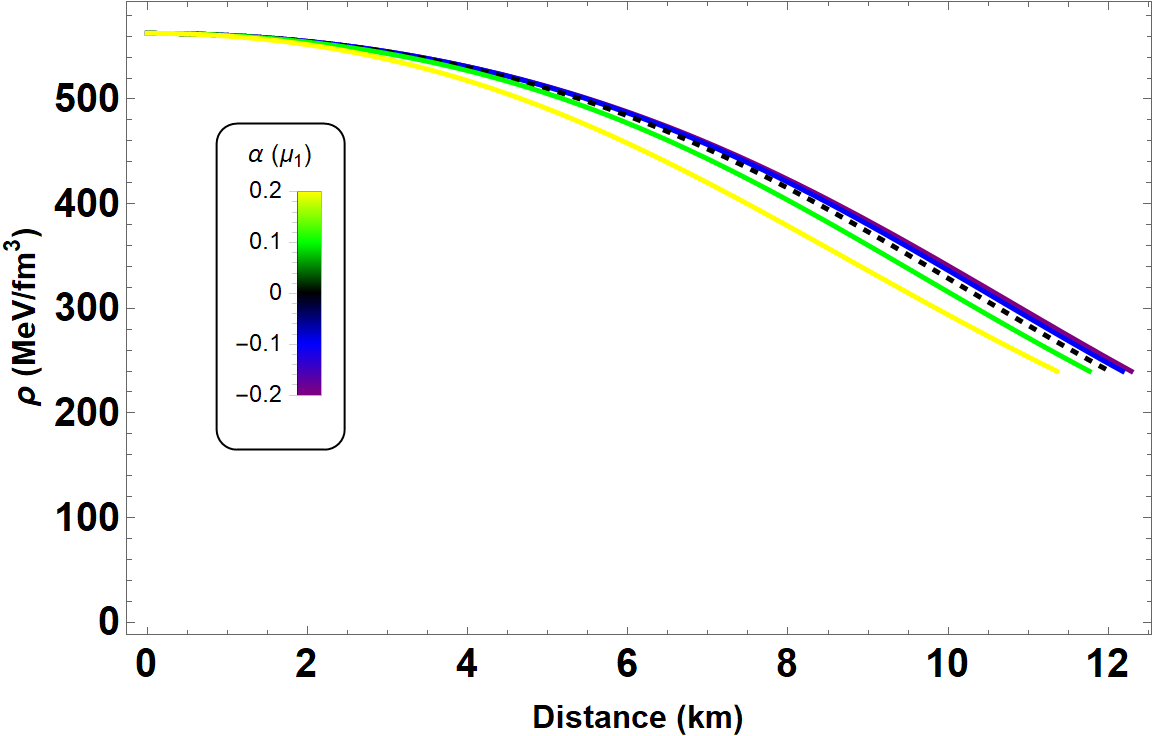}
    \includegraphics[width = 8 cm]{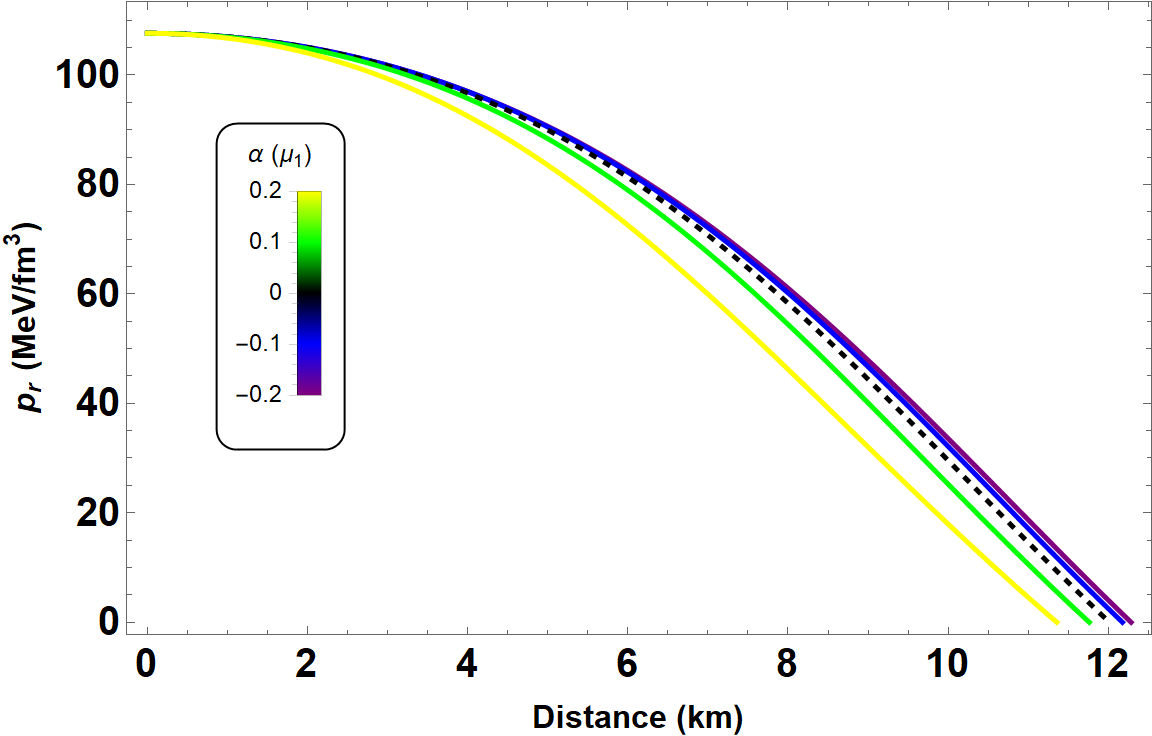}
    \includegraphics[width = 8 cm]{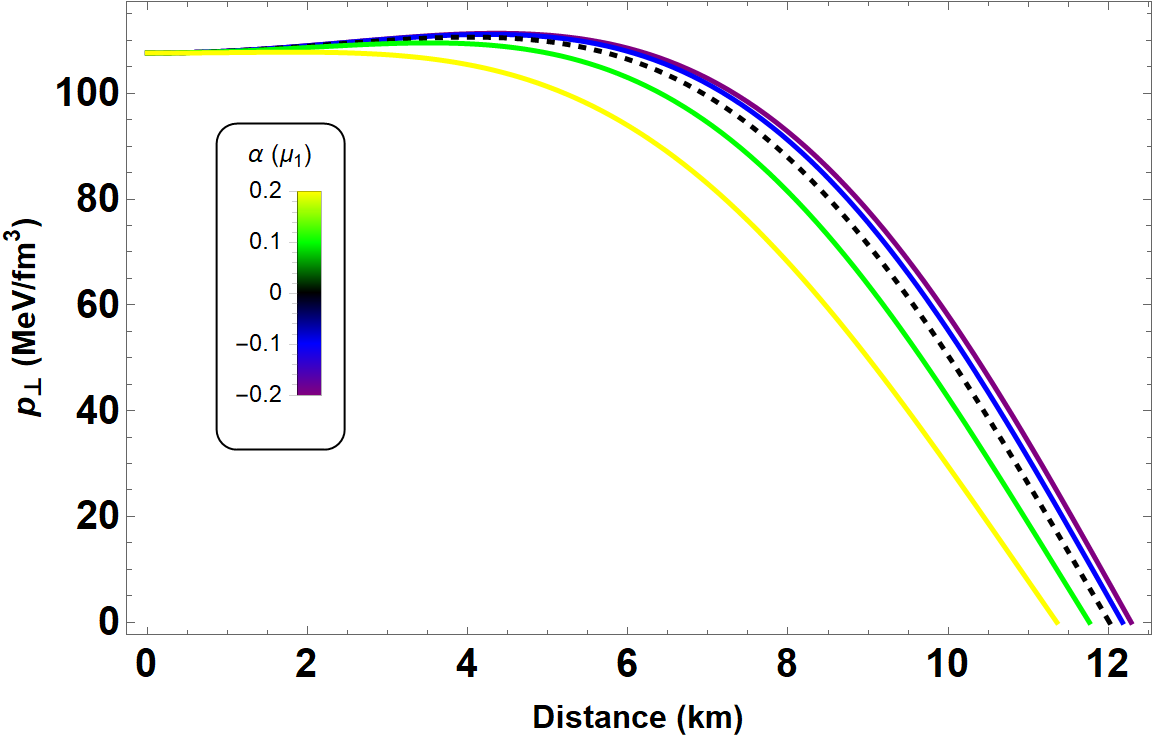}
\caption{From top to bottom, we display energy density $\rho$, radial $p_r$ and transverse  $p_t$ pressures as functions of the radial coordinate $r$. We start by varying $\alpha \in [-0.2,0.2] \mu_1$ where $\mu_1 = 10^{-79}$ s$^4$/kg$^2$. We define the other parameters are $B = 60$ MeV/fm$^3$ and $\beta = 1.5$, respectively. The dashed black line represents the GR solution $(\alpha = 0)$ for an anisotropic fluid distribution.}
   
    \label{fig1}
\end{figure}

\begin{figure}
    \centering
    \includegraphics[width = 8 cm]{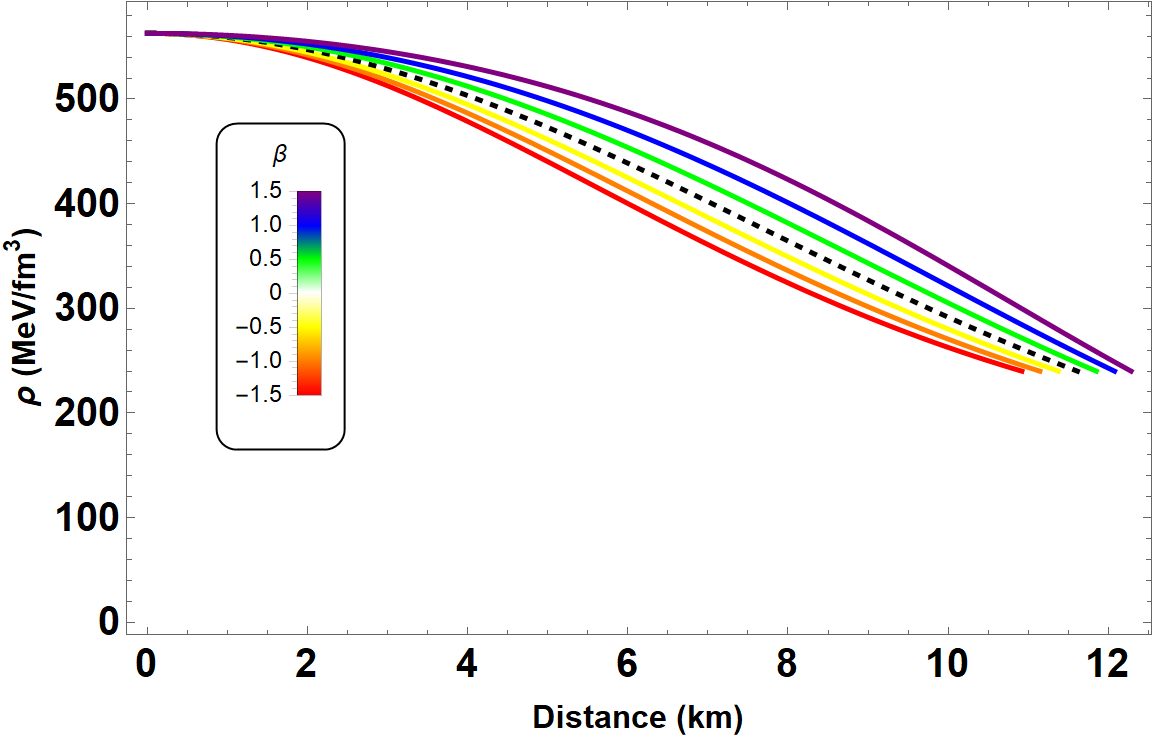}
    \includegraphics[width = 8 cm]{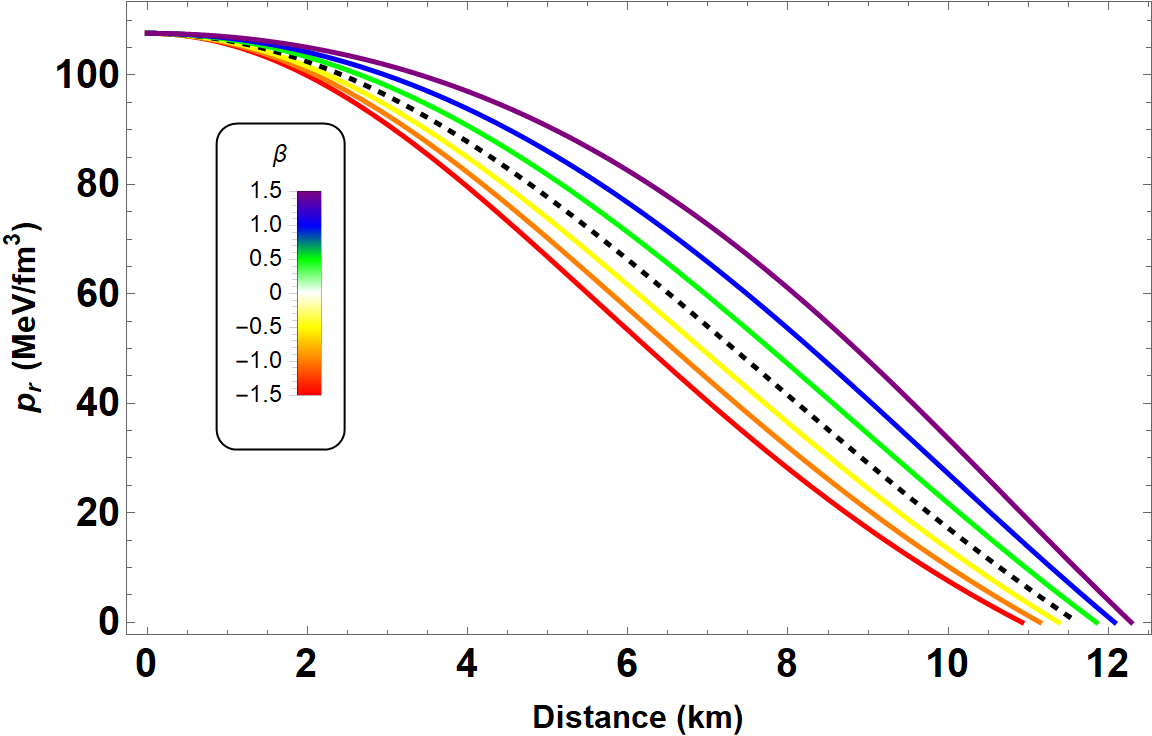}
    \includegraphics[width = 8 cm]{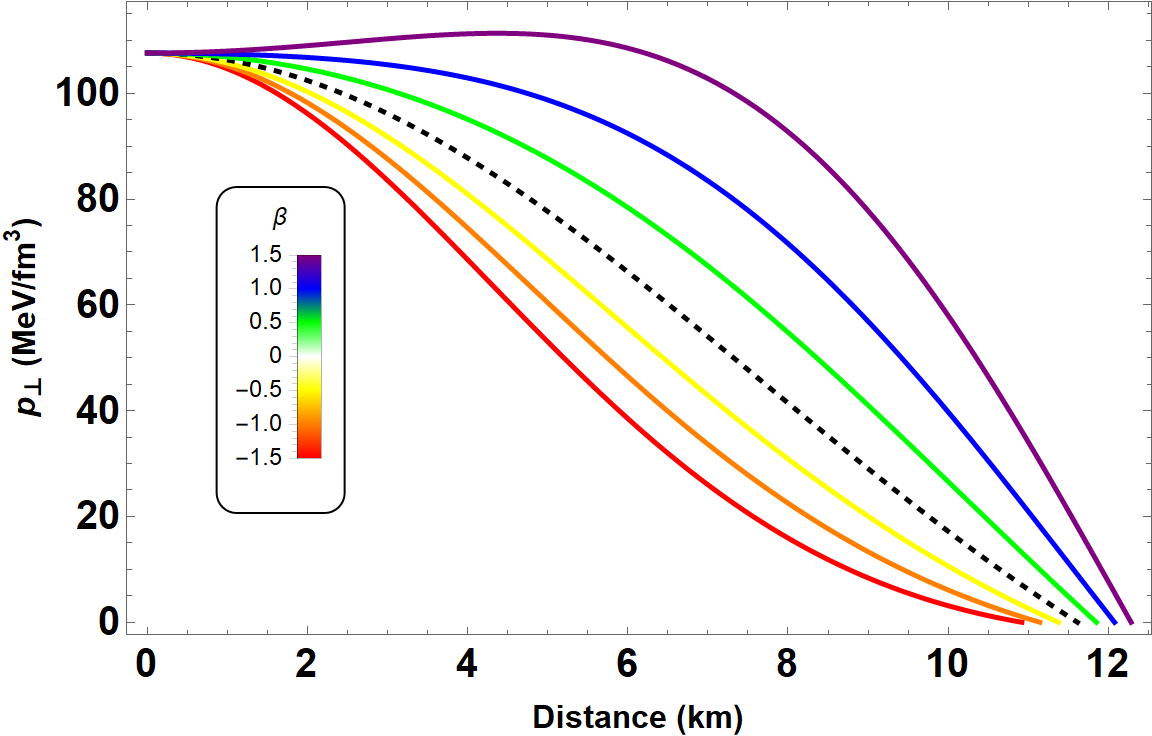}
     \caption{From top to bottom, we display energy density $\rho$, radial $p_r$ and transverse  $p_t$ pressures as functions of the radial coordinate $r$. We start by varying the anisotropy parameter $\beta \in [-1.5, 1.5]$, whereas the other parameters are $B = 60$ MeV/fm$^3$ and $\alpha =-0.2 \mu_1$ with $\mu_1 = 10^{-79}$ s$^4$/kg$^2$, respectively. The black dashed line denotes an isotropic arrangement in $f(R,L_m,T)$ gravity.}
    \label{fig2}
\end{figure}

Figures \ref{fig1} and \ref{fig2} depict various physical quantities as functions of the radial coordinate $r$, illustrating the sensitivity of the system to parameters $\alpha$ and $\beta$ respectively.

Before delving into the internal composition of compact stars in $f(R,L_m,T) = R + \alpha T L_m$ gravity, we introduce a mass function $m(r)$ through the relation:
\begin{equation} \label{eq18}
e^{-\lambda(r)} = 1 - \frac{2m(r)}{r},
\end{equation}
where $m(r)$ represents the total mass enclosed within the radius $r$. Subsequently, by substituting this expression into equations (\ref{FERho00})-(\ref{eq21}), we ultimately derive:
\begin{align}
    \frac{dm}{dr} &= 4\pi r^2\rho + \frac{\alpha r^2}{4}(p_r + 2 p_{\perp} - \rho)\rho ,  \label{eq22}  \\
    \frac{dp}{dr} &= -\frac{\Tilde{p}}{\Tilde{p}+ \alpha p_r} \bigg( \frac{(p_r + \rho)}{4r (r- 2m)}(4m + r^3 (p_r \Tilde{p} - 2 \alpha \rho^2) ) \nonumber \\
    &\quad  +\frac{2 \Delta}{r} + \frac{1}{\Tilde{p}} \left[ \alpha \{ p_r (2 p_{\perp}' - 3 \rho') - 4 \alpha \rho \rho' \} \right] \bigg). \label{eq23}
\end{align}

In this framework, we encounter five unknown functions with three differential equations. However, by introducing two appropriate ad hoc assumptions, we can close the system of equations. In this context, we consider an equation of state (EoS) relating radial pressure to the energy density of the fluid and a quasi-local EoS proposed by Horvat \textit{et al.} in Ref. \cite{Horvat:2010xf}, \textcolor{black}{which offeres reliable solutions and aligned effectively with GW170817 data.} With these considerations, we can numerically solve the structure equations with the appropriate boundary conditions:
\begin{align} \label{boundary}
m(0) &= 0, & \rho(0) &= \rho_c ,
\end{align}
ensuring regularity at the center of the star, where $\rho_c$ denotes the central energy density. Finally, we solve these equations by integrating outward until the pressure vanishes, i.e., $p_r(R) = p_{\perp}(R) = 0$, where $R$ is identified as the radius of the star. At this juncture, the interior solution is matched with the exterior Schwarzschild vacuum solution:

\begin{eqnarray}
e^{\nu(R)} = e^{-\lambda(R)} = 1 - \frac{2M}{R},
\end{eqnarray}

where $M$ signifies the total mass of the star.

\section{Equation of state for quark matter and the anisotropy  profile} \label{sec4}


Observations of compact objects via gravitational-wave (GW) signals or other recent astrophysical data have provided strong and direct evidence in favor of more massive compact stars. Interestingly, these observations also impose constraints on the NS EoS, suggesting that their internal composition likely includes free quarks in their core rather than solely neutrons and other non-fundamental particles. 
The central density is expected to reach several times higher than the nuclear saturation density ($n_0 \approx 0.148\ \text{fm}^{-3}$), leading to the possible formation of quark matter \cite{Hebeler:2010jx,Essick:2021ezp}. This density is typical within atomic nuclei and provides a reference point for understanding the extreme conditions that may lead to the deconfinement of quarks. This highly compressed nuclear matter is anticipated to undergo a phase transition to deconfined quark matter, releasing its constituent quarks and gluons. Therefore, the EoS of quark matter plays a pivotal role in shaping the structure of the star beyond nuclear saturation density. 
From Sec. \ref{sec4} onward, we now use "NS" units ($\hbar=c=1$) consistently in discussing the numerical results and the value of $\alpha$. The value of $\alpha$ mediated by $\mu_1$ is expressed using units appropriate for NS contexts. This ensures that all physical quantities are presented coherently throughout the manuscript.

In this work, we focus on the applications of strange matter, which are described by the MIT bag model EoS for modeling compact stellar objects. The MIT bag model is a simple phenomenological model for quark matter proposed in the 1970s to explain hadrons in terms of quarks \cite{Chodos:1974}. In this model, it is assumed that quarks are asymptotically free and confined to a spherical region of space by the bag constant $B$. Consequently, the MIT bag model establishes a relation between energy density and pressure, given by:
\begin{eqnarray}\label{Prad1}
    p_r = \dfrac{1}{3}\left(\rho -4B\right).
\end{eqnarray}

This expression indicates that at $\rho = 4B$, the external pressure acting on a bag vanishes. The constant $B$, referred to as the bag pressure, is typically constrained within a range of $57 \leq B \leq 92$ MeV/fm$^3$. This range is supported by various studies in the field \cite{Burgio:2018mcr, Blaschke:2018mqw}. Following \cite{Arbanil:2016wud}, we use $B = 60\ \rm MeV/fm^3$ to align the model with observational astrophysical data. Moreover, it is noteworthy that this chosen value also satisfies the $2M_\odot$ constraints within the framework of GR. 
For local anisotropy, we employed the quasilocal EoS suggested by Horvat et al. \cite{Horvat:2010xf}. The choice of this model was driven by two key factors: (a) The Horvat model's simplicity and its physical consistency, which connects the anisotropy directly to the local compactness parameter $\mu = \frac{2m(r)}{r}$, making it an intuitive choice for compact stars. (b) Furthermore, as shown in \cite{isMota:2024kjb}, the Horvat model emerges naturally from the modified TOV equation for $dp_r/dr$, making it a suitable and physically grounded approach for studying anisotropic compact stars. While other anisotropic models could have been considered, such as those based on pressure or density gradients, the Horvat model offers a straightforward and analytically tractable way to explore anisotropic effects. This model's ability to smoothly transition between isotropic and anisotropic cases by varying the parameter $\beta$ further supports its use in our analysis. Within this framework, various solutions have been obtained in both GR and modified gravity theory. Thus, we express the quasi-local equation as:
\begin{eqnarray}\label{anisotropy}
    \Delta &\equiv& p_{\perp} - p_r =  \beta p_r \mu,
\end{eqnarray}
where the free parameter $\beta$ plays a crucial role in measuring the deviation from isotropy, taking positive or negative values, see Refs. \cite{Doneva:2012rd, Silva:2014fca, Yagi:2015hda, Pretel:2020xuo, Rahmansyah:2020gar, Rahmansyah:2021gzt, Folomeev:2018ioy} for detailed discussions. In our calculations, we assume $\beta \in [-1.5,1.5]$. The quantity $\mu \equiv 2m(r)/r$ represents the local measure of compactness. Furthermore, the choice of Eq. (\ref{anisotropy}) ensures that $\Delta =0$ at the center, thereby recovering the isotropic case. In Figs.~\ref{fig1} and \ref{fig2}, we depict the energy density $\rho$, radial pressure $p_r$, and transverse pressure $p_t$ as functions of the radial coordinate $r$ for two different parameter sets (see Tables \ref{tableVaryalpha} and \ref{tableVarybeta} for more details). It is evident from these figures that the energy density of the QS is non-zero at the surface, while the pressure components vanish simultaneously at the surface of the star, i.e., $p_r \left( r \rightarrow R \right) = p_{\perp} \left( r \rightarrow R \right) = 0$. Clearly, when $\beta =0$, the anisotropy factor vanishes at the center, ensuring regularity in the interior.


\section{Numerical results and discussion}\label{sec5}

To investigate the influence of pressure anisotropy on the stellar structure,  we first integrate numerically the equations of  \eqref{eq22}, \eqref{eq23}, \eqref{Prad1} and \eqref{anisotropy} together with the boundary conditions \eqref{boundary}.  By varying the central density, we obtain mass as a functions of the radial coordinate $r$, the mass-radius ($M-R$) and the compactness $(M-M/R)$ relations. Here, we use several values of $\alpha \in [-0.2,0.2] \mu_1$ where $\mu_1 = 10^{-79}$ s$^4$/kg$^2$ \cite{isMota:2024kjb} and $\beta \in [-1.5, 1.5]$ to demonstrate the results for anisotropic stars in $f(R,L_m,T)= R+ \alpha T L_m$ gravity.  

\begin{figure}
    \centering
    \includegraphics[width = 8 cm]{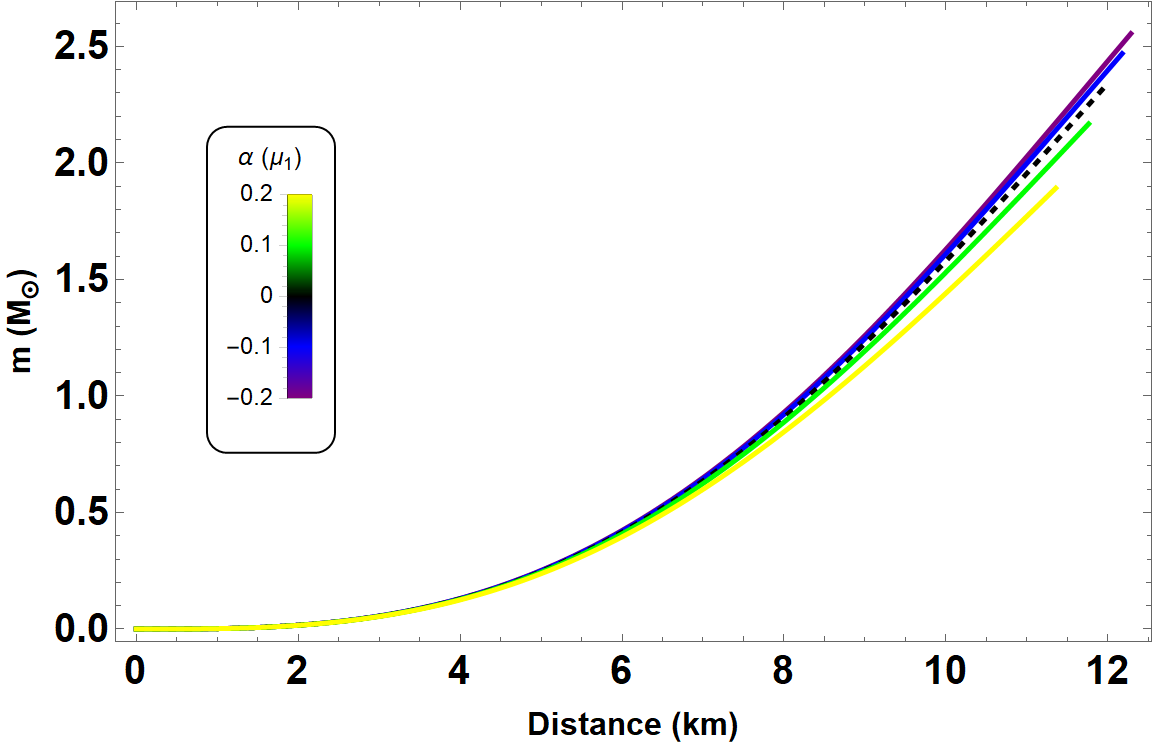}
    \includegraphics[width = 8 cm]{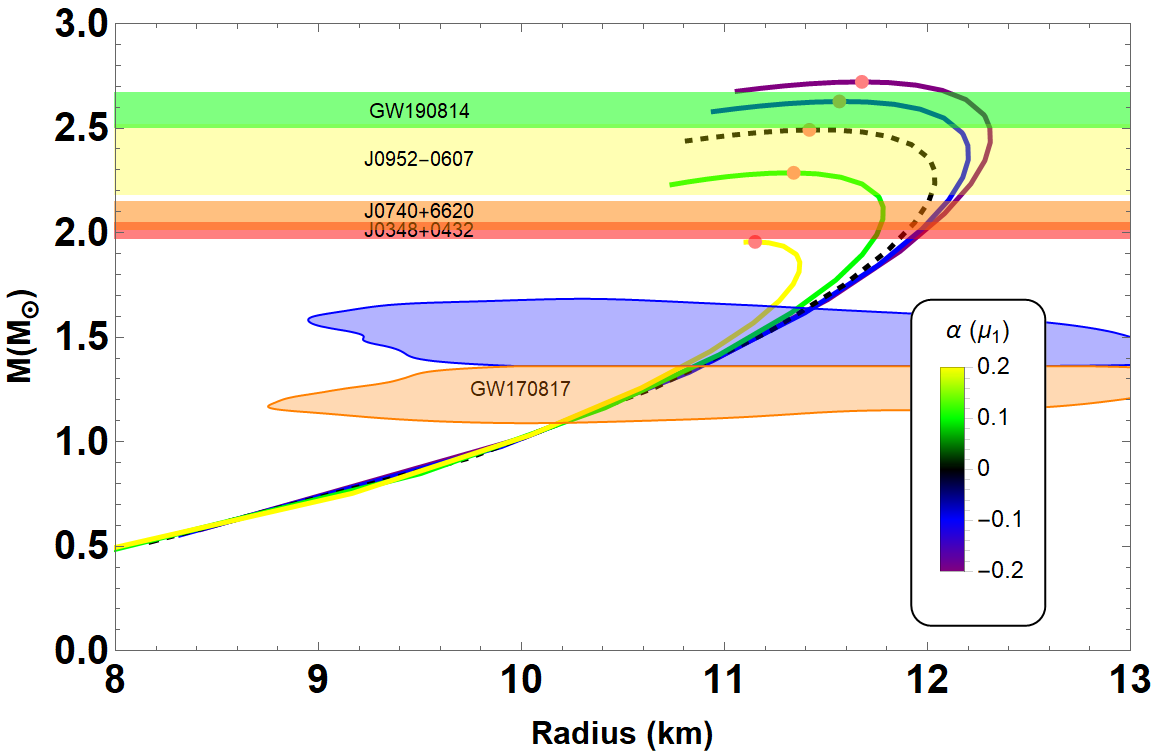}
    \includegraphics[width = 8 cm]{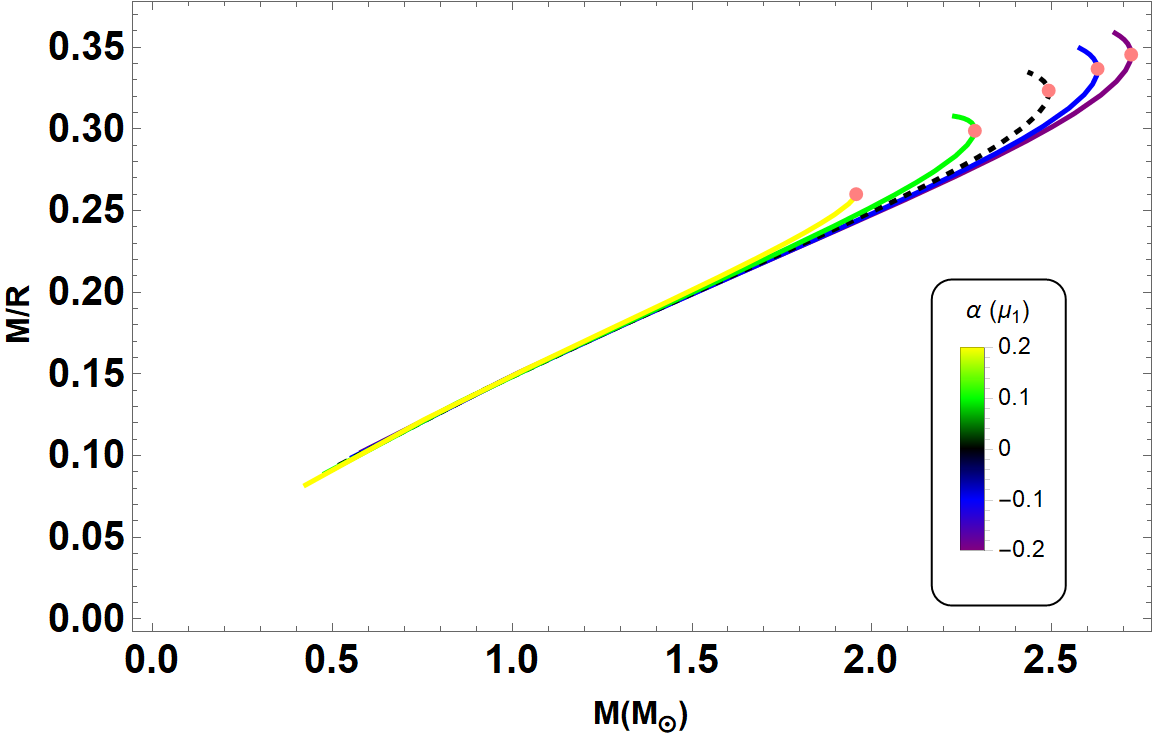}
    \caption{The mass as a function of radial coordinate (top), the mass-radius (middle), the mass-compactness (bottom) relations for MIT bag model EoS used in this paper with parameters mentioned in Table \ref{tableVaryalpha}. The horizontal bands correspond to the observed masses of pulsars:  PSR J0952-0607 \cite{Romani:2022jhd}, PSR J0740+6620  \cite{Fonseca:2021wxt}  and PSR J0348+0432  \cite{Antoniadis:2013pzd} with different colour bars. Furthermore, we include the constraint from GW190814 \cite{LIGOScientific:2020zkf} and GW170817 event \cite{LIGOScientific:2018cki}, also. In the relevant graphs, the black dashed line represents the GR solution $(\alpha = 0)$ for an anisotropic fluid distribution.}
    \label{fig_vary_alpha}
\end{figure}

\begin{figure}
    \centering
    \includegraphics[width = 8 cm]{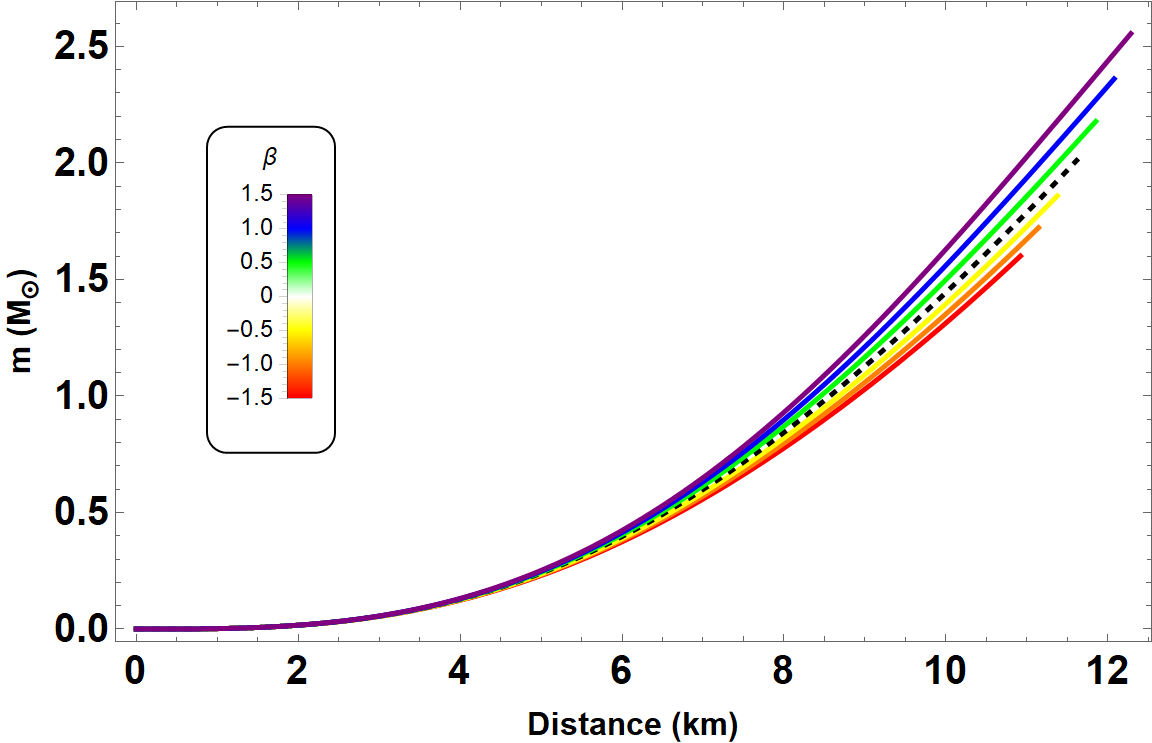}
    \includegraphics[width = 8 cm]{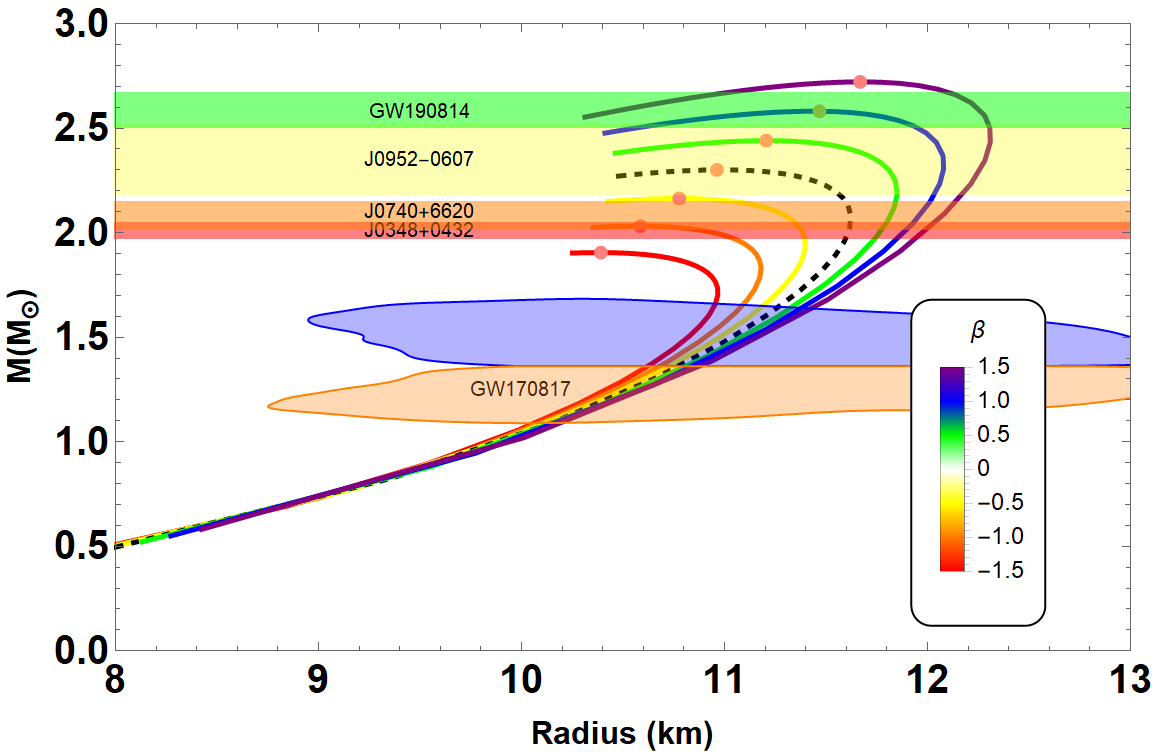}
    \includegraphics[width = 8 cm]{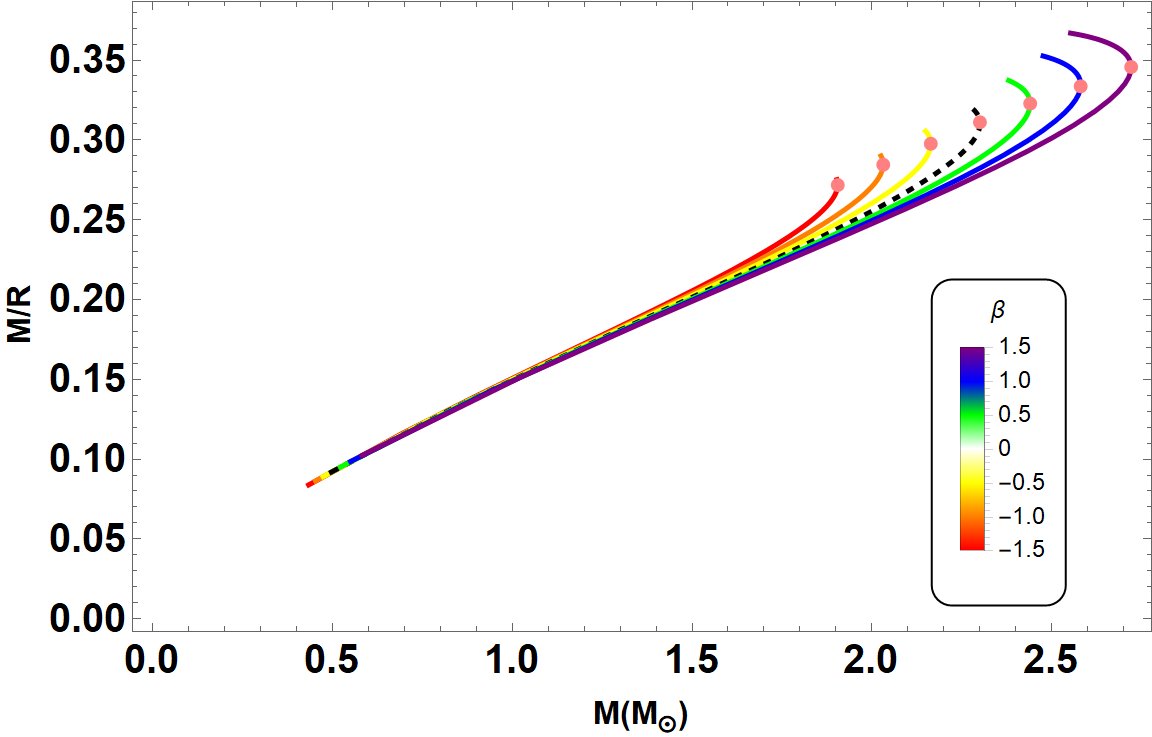}
    \caption{The mass as a function of radial coordinate (top), the mass-radius (middle), the mass-compactness (bottom) relations for MIT bag model EoS used in this paper with parameters mentioned in Table \ref{tableVarybeta}. Imposing constraints from astrophysical observations are the same as of Fig. \ref{fig2}. In the relevant graphs, the black dashed line represents an isotropic configuration in $f(R,L_m,T)$ gravity.}
    \label{fig_vary_beta}
\end{figure}

\subsection{Profiles for varying $\alpha$}\label{case1}

From top to bottom, we plot $M-r$, $M-R$ and $(M/R-M)$ curves for the variation of $\alpha \in [-0.2,0.2] \mu_1$ where $\mu_1 = 10^{-79}$ s$^4$/kg$^2$ in Fig. \ref{fig_vary_alpha}. The other parameters are $B = 60$ MeV/fm$^3$ and $\beta = 1.5$, respectively. For this particular model and data set, we found that the mass of the QS increases for decreasing values $\alpha$. This observation is consistent with the results reported in Ref. \cite{isMota:2024kjb}. Indeed, the reported value of maximum mass for QS is 2.72 $M_{\odot}$ with radius $R=11.68$ km for $\alpha = -0.2 \mu_1$. For completeness, we list the maximum masses and corresponding radii of QSs obtained from the numerical calculation in Table \ref{tableVaryalpha} from where we can quantify how coupling constant $\alpha$ is contributing to $M_{\rm{max}}$ estimates. Moreover, the maximum mass in GR (for anisotropic fluid) is 2.49 $M_{\odot}$ with radius $R=11.42$ km. In other words, we can say that sufficiently negative values of $\alpha$ increase the maximum mass and the corresponding radius also, see Table \ref{tableVaryalpha}. It is worth noting that the deviation between GR and $f(R,L_m,T)$ gravity is significant. Putting the observational constraints into consideration, we consider several observations, in particular, PSR J0952-0607 with mass $M = 2.35 \pm 0.17 M_{\odot}$ (Yellow) \cite{Romani:2022jhd}, PSR J0740+6620 with the pulsar mass $ M = 2.08 \pm 0.07$ $M_{\odot}$ (Orange) \cite{Fonseca:2021wxt} and PSR J0348+0432 with the mass of $ M = 2.01 \pm 0.04 M_{\odot}$ (Pink) \cite{Antoniadis:2013pzd}. We can also discuss our results in light of constraints obtained from the detection of GW190814 event with a mass of $2.59^{+0.08}_{-0.09} M_{\odot}$ (Green) \cite{LIGOScientific:2020zkf} and GW170817 event \cite{LIGOScientific:2018cki}, see middle panel of Fig. \ref{fig_vary_alpha}. More importantly, we can say that our results are consistent with the main constraint from recent observational astrophysical data. Finally,  we plot $(M-M/R)$ diagram in the lower panel of Fig. \ref{fig_vary_alpha}. We see that for different values of $\alpha$ the maximum compactness is different corresponding to the maximum masses. One also notes from Table \ref{tableVaryalpha} that the value of maximum compactness increases with decreasing values of $\alpha$, which is consistent with the $M-R$ diagram.  Moreover, the  value of maximum compactness lies within the range of  $0.260 < M/R < 0.345$. \textcolor{black}{It is also worth noting that theoretical considerations \cite{isAstashenok:2014pua,isAndersson:2020phh} suggest \( \alpha \) should be small to avoid large deviations from GR predictions, especially for stable, high-density QS configurations. Observationally, fitting to specific QS data could provide indirect constraints. Large negative \( \alpha \) enhances mass, while positive \( \alpha \) tends to reduce it, as shown in Table \ref{tableVaryalpha}. These values influence compactness and stability within observed astrophysical ranges.}


\begin{table}[h]
    \caption{ Summary of the resulting properties of anisotropic QSs for the variation of $\alpha$ in $f(R,L_m,T)$ gravity.  A brief discussion is given in Sec. \ref{case1}.}
    \begin{ruledtabular}
    \begin{tabular}{ccccc}
    $\alpha$  & $M$ & $R$ & $\rho_c$ & $M/R$\\
    $\times 10^{-78} \frac{s^4}{kg^2}$  & $M_{\odot}$ &   km & MeV/fm$^3$ & \\
    \hline
        -0.2 &  2.72  & 11.68 & 847 & 0.345 \\
        -0.1 &  2.63 & 11.57 & 847 & 0.337 \\
        0.0 &  2.49  & 11.42 & 847 & 0.323 \\
        0.1 &  2.29  & 11.34 & 791 & 0.299 \\
        0.2 &  1.96  & 11.15 & 735 & 0.260 
    \end{tabular}
    \end{ruledtabular}
    \label{tableVaryalpha}
\end{table}

\subsection{Profiles for varying $\beta$}\label{case2}

We further analyze the effect of anisotropic parameter $\beta \in [-1.5, 1.5]$ on the stellar structure in Fig. \ref{fig_vary_beta}. The sequence of curves that describe the behavior of $M-r$, $M-R$ and $(M/R-M)$ for the parameter set: $B = 60$ MeV/fm$^3$ and $\alpha = -0.2 \times 10^{-78} s^4/kg^2$. From the figure, we see that the mass of the QS increases for increasing values $\beta$. The result of the main aspects are summarized in  Table \ref{tableVarybeta}, and Fig. \ref{fig_vary_beta} (middle panel) shows some associated $M-R$ relations for QSs. Depending on the model parameter, the value of maximum mass is $M_{\rm{max}}=$ 2.72 $M_{\odot}$ and the corresponding radius is $11.67$ km at $\beta = 1.5$. As for the values of $\beta =0$ (the isotropic fluid) the maximum mass goes upto $M_{\rm{max}}=$ 2.30 $M_{\odot}$ and the corresponding radius is $10.96$ km. As expected the presence of  pressure anisotropy leads to a considerable changes in the $M-R$ relations. Concerning the constraints, we see that our model is compatible with maximum QS masses. At the end of this discussion, the $(M-M/R)$ diagram has been plotted in Fig. \ref{fig_vary_beta} (lower panel). The  maximum compactness of QSs is monotonically increases with increasing $\beta$ values. Finally, the results are presented in Table \ref{tableVarybeta}, where we can see that the  maximum compactness corresponding to the maximum mass lies within the range of  $0.272 < M/R < 0.346$. 


\begin{table}
    \caption{Summary of the resulting properties of anisotropic QSs for the variation of $\beta$ in $f(R,L_m,T)$ gravity. A brief discussion is given in Sec. \ref{case2}. }
    \begin{ruledtabular}
    \begin{tabular}{ccccc}
    $\beta$  & $M$ & $R$ & $\rho_c$ & $M/R$\\
    & $M_{\odot}$ &   km & MeV/fm$^3$ & \\
    \hline
    -1.5 & 1.91 & 10.39 & 1,415 & 0.272 \\
    -1.0 & 2.03 & 10.59 & 1,334 & 0.284 \\
    -0.5 & 2.16 & 10.78 & 1,253 & 0.298 \\
    0.0 & 2.30 & 10.96 & 1,173 & 0.311\\
    0.5 & 2.44 & 11.20 & 1,052 & 0.323\\
    1.0 & 2.58 & 11.47 & 931 & 0.334\\
    1.5 & 2.72 & 11.67 & 851 & 0.346\\
    \end{tabular}
    \end{ruledtabular}
    \label{tableVarybeta}
\end{table}

\section{The static stability criterion, adiabatic index and the sound velocity}\label{sec6}

Apart from the  mass-radius relations, the most important issue is related to stability of the configuration. Below we 
start by studying in details. 

\begin{figure}[h]
 \centering
 \includegraphics[width = 8 cm]{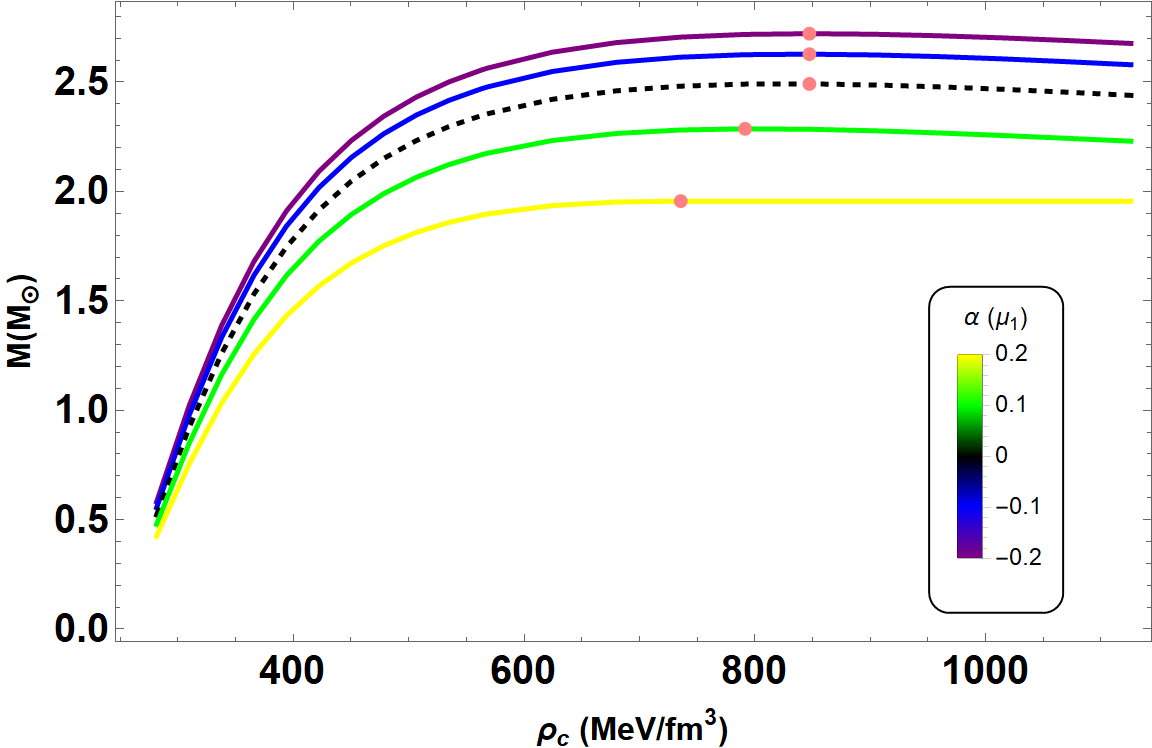}
  \includegraphics[width = 8 cm]{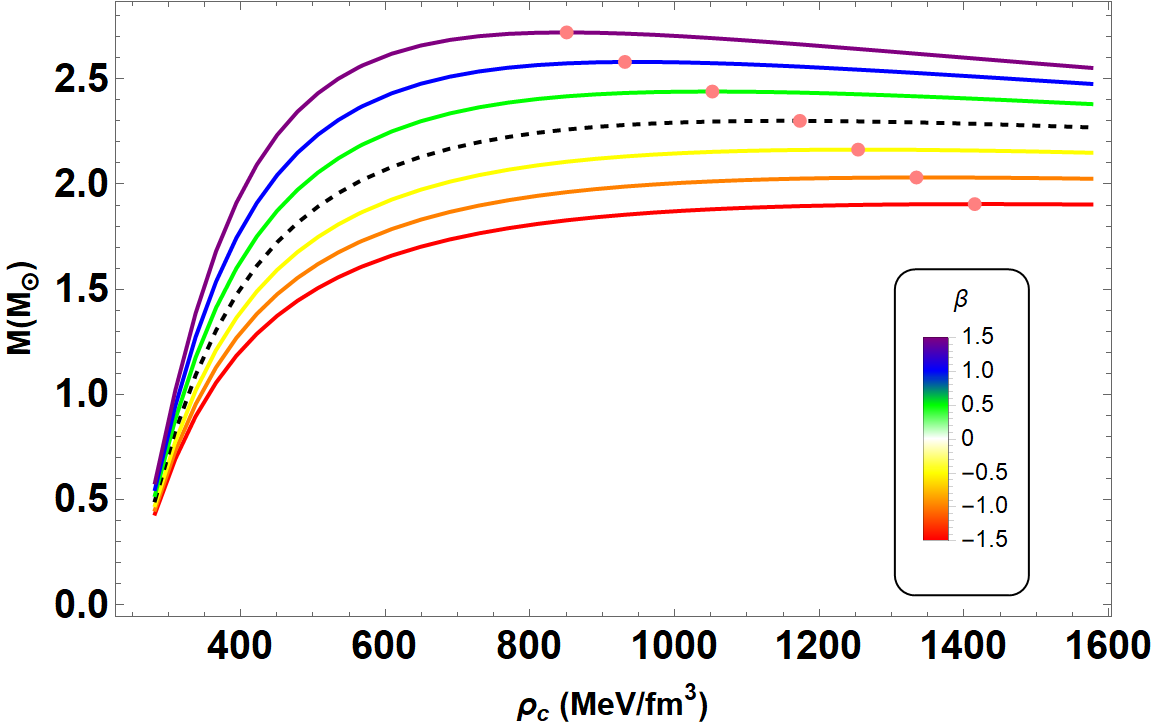}
 \caption{This profile is for $M-\rho_c$ curves. The same parameter set as used for Figs. \ref{fig1} to \ref{fig2} is considered. The pink points are representing as a boundary wall that separates the stable configuration region indicated by $dM/d\rho_c >0$ from the unstable one.} \label{fig_SSC}
 \end{figure}

\subsection{Static stability criterion}

Let us now focus on the stability of the equilibrium configuration through \textit{static stability criterion} (SSC) \cite{harrison,ZN}. This criteria has been presented in  $M-\rho_c$ plane,  where $M$  is the mass and $\rho_c$ is the central density of the star. In a similar vein, we see that SSC has also been applied in modified gravity theory also, see Refs. \cite{Maulana:2019sgd,Pretel:2022plg,Gammon:2023uss} for more. We should also note that this is a necessary but insufficient condition. The mathematical form of these inequalities is as follows 
\begin{eqnarray}
\frac{d M}{d \rho_c} < 0 &~ \rightarrow \text{indicating an unstable configuration}, \\
\frac{d M}{d \rho_c} > 0 &~ \rightarrow \text{indicating a stable configuration}.
\label{criterion_M_rho_c}
\end{eqnarray}
In Fig. \ref{fig_SSC}, we show the $M-\rho_c$ curves for the consider cases mentioned above. For plotting we use the same parameters as shown in Fig. \ref{fig1} and \ref{fig2}, respectively. Here, we see that the total mass is
an increasing function of the central density, and reach at a  point where $(M_{\text{max}}, R_{M_{\text{max}}})$ exists. This point is known as a boundary point (indicated in the figure by pink points) which can separate the stable configuration region indicated by $dM/d\rho_c >0$ from the unstable one.

\subsection{Adiabatic indices}

\begin{figure}[h]
    \centering
    \includegraphics[width = 8 cm]{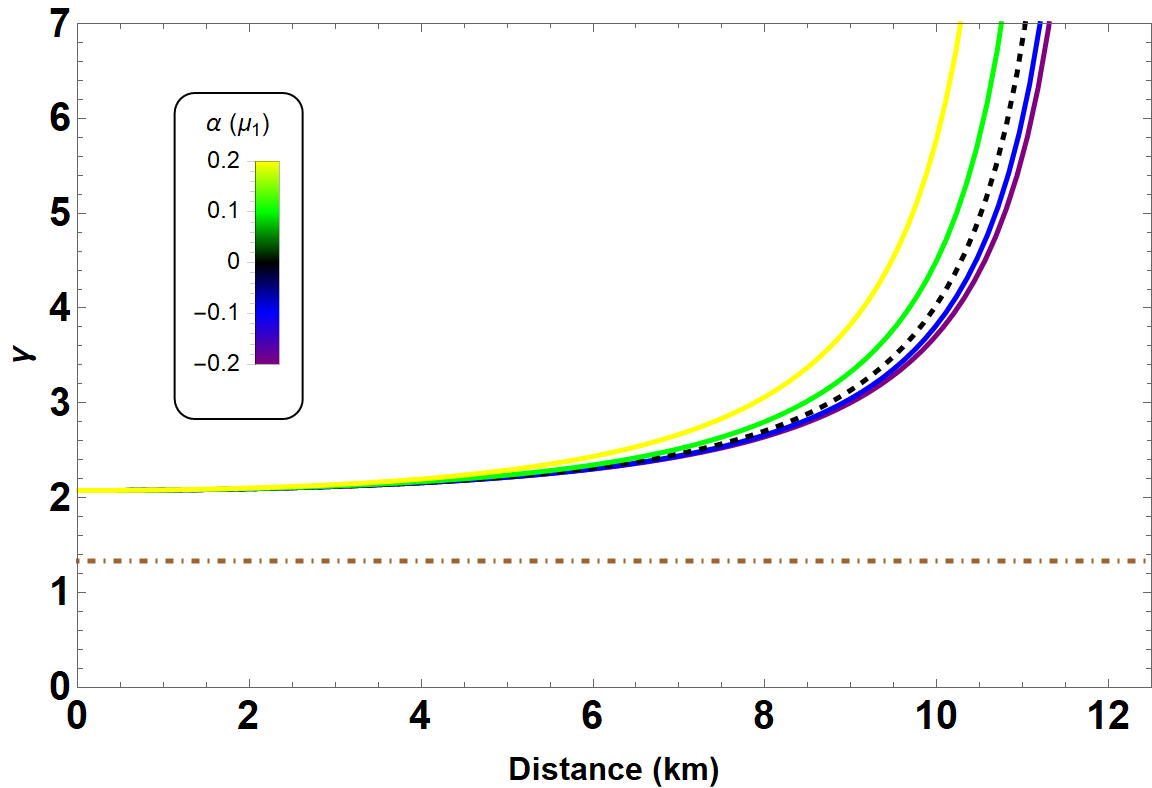}
    \includegraphics[width = 8 cm]{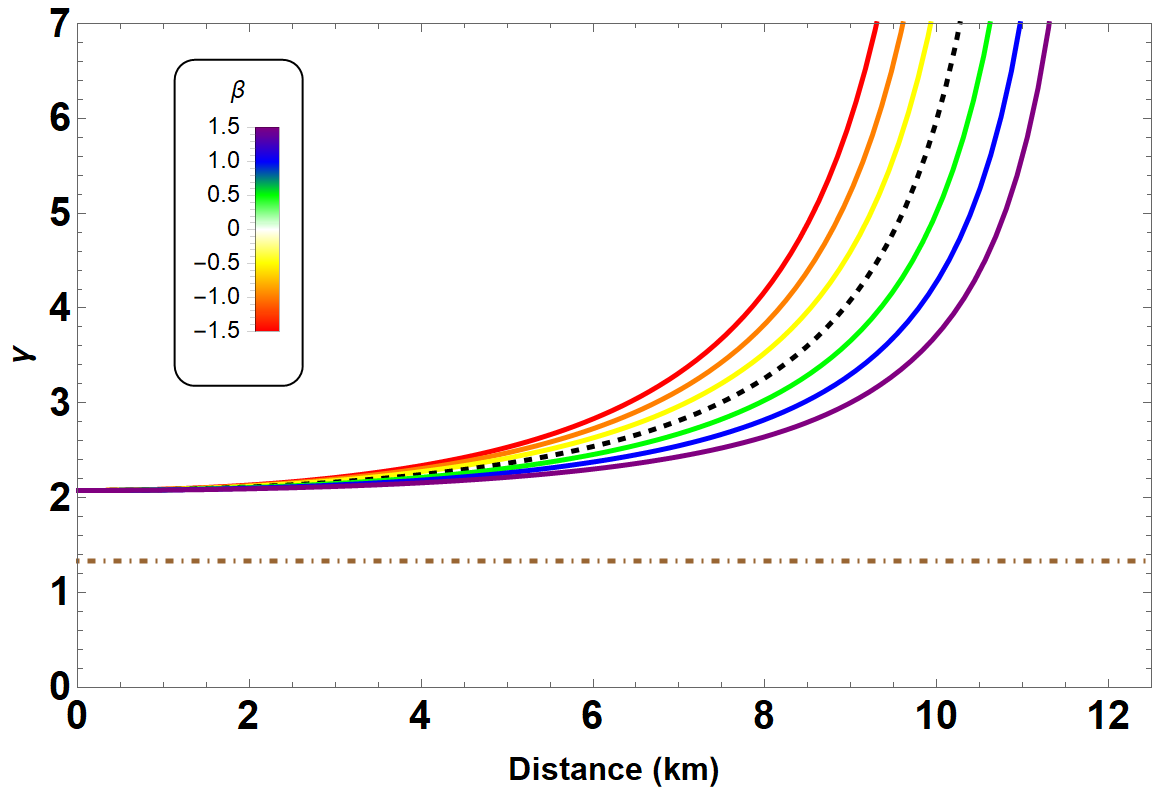}
    \caption{The adiabatic index $\gamma$ is plotted as a function of radial coordinate $r$. The same parameter set as used for Figs. \ref{fig1} and \ref{fig2} is considered. In figures we indicate the value of  $\gamma_{cr}= 4/3$ by the brown dot-dash line. }
    \label{fig_adiabatic}
\end{figure}

An additional test has been performed to check the stability of the configuration via adiabatic index, $\gamma$. The stability of compact objects was extensively studied, with notable contributions by Chandrasekhar \cite{Chandrasekhar:1964zz,Chandrasekhar:1964zza}, who laid the groundwork for understanding dynamical instability in compact stars. His work provides key insights into the limits of stability for stellar objects, including NSs and white dwarfs. Keeping the spherical symmetry of the background, Chandrasekhar performed the dynamical stability using the theory of infinitesimal radial perturbations \cite{Chandrasekhar:1964zz}. The adiabatic index ($\gamma$) is given by   
\begin{eqnarray}\label{adi}
    \gamma \equiv \left(1+\frac{\rho}{p_r}\right)\left(\frac{dp_r}{d\rho}\right)_S. 
\end{eqnarray}
The expression \eqref{adi} is associated with the sound speed and the subscript $S$ indicates the derivation at constant entropy.  It's also worth mentioning that $\gamma$ has some restrictions which is related with the dynamical instability of the spherical static object. For polytropic stellar model, authors in \cite{Glass} have shown that the adiabatic index $\gamma$ is greater than $\gamma> \gamma_{cr} = 4/3$, below which configurations are unstable against radial perturbations.
For more details about the role of adiabatic index on stellar structure see Ref. \cite{Moustakidis:2016ndw}. In Fig. \ref{fig_adiabatic}, we display the dependence of $\gamma$ for two considered cases as a function of radial coordinate $r$. The figures lead to the conclusion that $\gamma > \gamma_{cr}$ holds for both cases, ensuring the stability of the configuration under consideration. 

\subsection{Sound speed and  causality} 

\begin{figure}
    \centering
    \includegraphics[width = 8 cm]{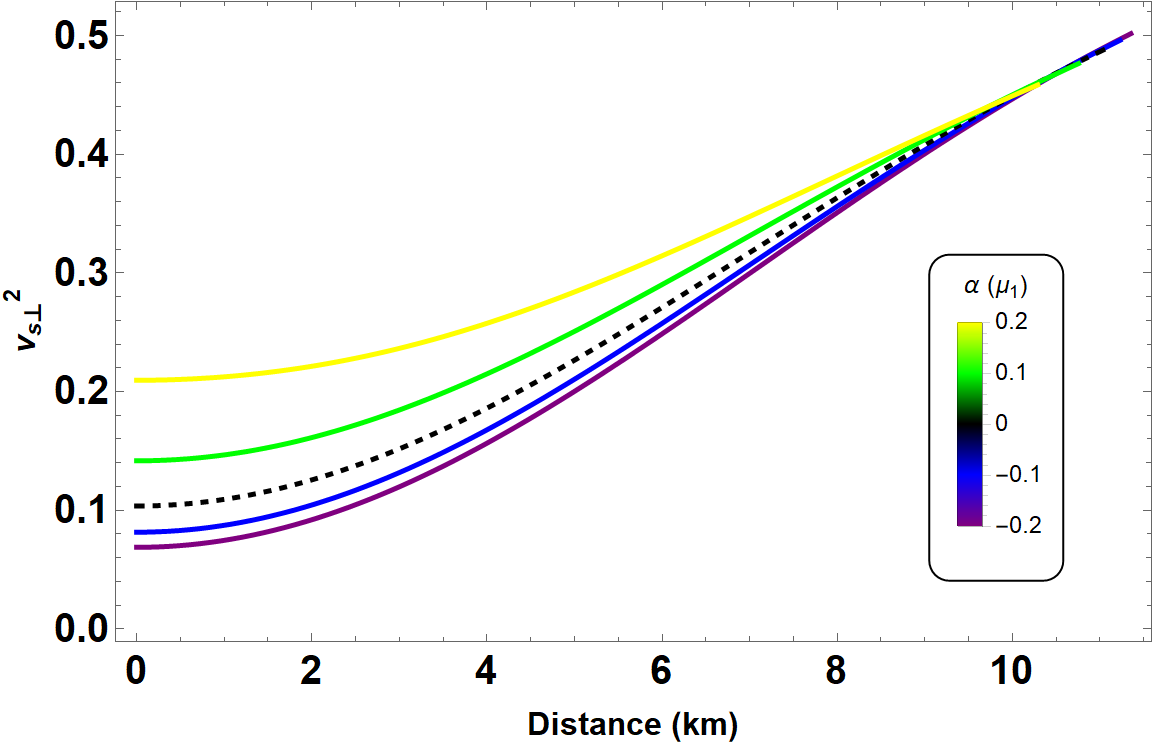}
    \includegraphics[width = 8 cm]{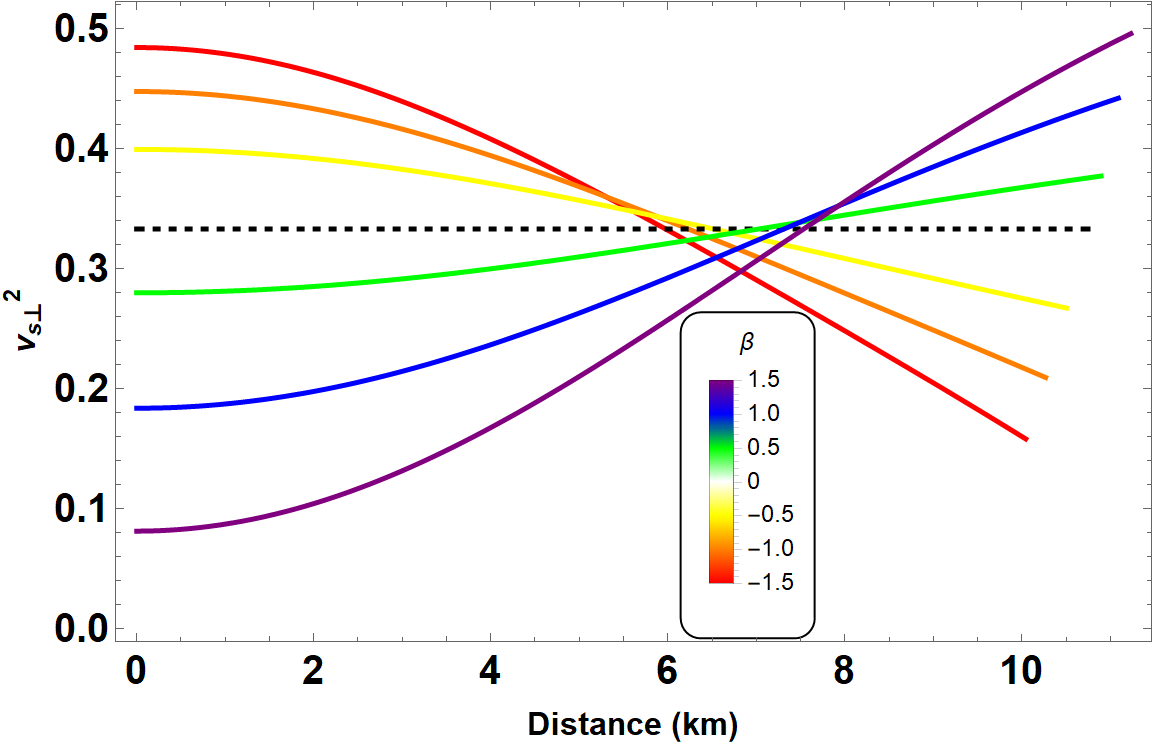}
    \caption{This plot is for sound speed along transverse direction within the stellar interior.  
    The same parameter set as used for Figs. \ref{fig1} and \ref{fig2} is considered.}
    \label{fig_soundspeed}
\end{figure}

The propagation of sound speed is another important criteria for checking the stability of QSs. The sound speed within the star is given by $ v^2_s = dp_{r, \perp}/ d\rho$, and lies within the range of $ 0<v^2_s <1$. Because the speed of sound does not exceed the speed of light. Since, the sound speed along radial direction is constant inside the star, whereas the tangential velocity has been plotted (both cases) in
Fig. \ref{fig_soundspeed}. From those figures, we see that the sound speed along transverse direction lies within the specified range. Thus,  we can say that 
QSs with anisotropic pressure could exists in $f(R,L_m,T)$ gravity.


\section{ concluding remarks }\label{sec7}
In this study, we investigated the behavior of QSs within the framework of $f(R,L_m,T)$ gravity, focusing on the specific model $f(R,L_m,T) = R + \alpha T L_m$. This model offers a novel perspective on the internal structure of compact stars, allowing us to explore the impact of modifications to the gravitational action on the properties of QSs.

Our analysis revealed several key findings. First, we derived the modified TOV equations for QSs in $f(R,L_m,T)$ gravity, considering a specific choice of the matter Lagrangian density $L_m= -\rho$. These equations govern the hydrostatic equilibrium of QSs and provide valuable insights into their macroscopic properties.

Next, the modified TOV equations are solved numerically with proper boundary conditions to demonstrate the results for anisotropic stars in $f(R,L_m,T)= R+ \alpha T L_m$ gravity. Our results showed
the sensitivity of QS properties to the MIT bag parameter, highlighting the importance of considering various scenarios when 
studying compact stars in modified gravity theories. This model has been carried out for varying the coupling parameter  $\alpha$
and anisotropic parameter $\beta$ to find the maximum mass of QSs. The results are interesting since the maximum mass is 
increasing for sufficiently negative values of $\alpha$, while the situation is reverse for increasing positive values of $\beta$. Furthermore, we analyzed the static stability criterion, adiabatic index, and sound velocity of QSs to assess their dynamical stability. Our calculations showed that QS models within the considered parameter ranges satisfy stability criteria, indicating the robustness of these configurations against radial oscillations and sound propagation.

\textcolor{black}{Our modified QS model showed consistency with observed high-mass compact stars, such as PSR J0740+6620, which had an estimated mass of $M = 2.08 \pm 0.07 \, M_{\odot}$ and radius constraints provided by NICER data \cite{isMiller:2021qha}. The influence of the coupling parameter $\alpha$ in the $f(R, L_m, T)$ gravity model supported the attainment of higher maximum masses in QS configurations, aligning with this observational range. Additionally, gravitational wave data, particularly from events GW190814 and GW170817, imposed stringent constraints on compact star models. The high secondary mass in GW190814 ($2.50-2.67 \, M_{\odot}$) presented a challenge for traditional neutron star EoS but aligned with the higher mass predictions of our modified model. By adjusting the coupling parameter, we also achieved QS configurations compatible with the tidal deformability limits from GW170817 \cite{isDe:2018uhw}, indicating that our model could satisfy these observational constraints. This agreement with astrophysical data underscored the potential of $f(R, L_m, T)$ gravity as a framework for understanding the properties of ultra-dense compact objects.}

Overall, our study contributes to the growing body of research on compact stars in alternative gravity theories. By exploring the behavior of QSs in $f(R,L_m,T)$ gravity, we deepen our understanding of the gravitational effects on extreme astrophysical objects. \textcolor{black}{In conclusion, the nonlinear term \( \alpha T L_m \) might influence additional QS characteristics, including compactness and rotation, through adjustments to the stress-energy tensor and equilibrium equations. This term might suggest surface effects like tension and impact the moment of inertia during rapid rotation, which we intend to explore further in a separate study.} 

Future investigations could extend this work to other modified gravity models and explore additional observational consequences for QSs, such as gravitational wave signatures and electromagnetic emissions. Moreover, it would be interesting to explore the implications of our findings for the formation and evolution of quark stars in diverse astrophysical environments. Additionally, incorporating magnetic fields or studying rotating quark stars could provide further insights into the complex dynamics of these exotic objects.

\begin{acknowledgments}
\textcolor{black}{The authors would like to thank the Editor abd Reviewer for their insightful comments and constructive suggestions, which have greatly enhanced the clarity and quality of this manuscript. Their efforts have significantly contributed to the improvement and refinement of our work.} T. T. was supported by Walailak University under the New Researcher Development scheme (Contract Number WU67268). A. Pradhan expresses gratitude to the IUCCA in Pune, India, for offering facilities under associateship programs. In addition, \.{I}.S. thanks T\"{U}B\.{I}TAK, ANKOS, and SCOAP3 for their contributions. T.T. and {I}.S. also appreciate COST Actions CA21106 and CA22113 for their networking support.
\end{acknowledgments}\

\bibliography{sample631}{}
\bibliographystyle{aasjournal}
\end{document}